\documentclass[sigconf]{acmart}

\AtBeginDocument{%
  \providecommand\BibTeX{{%
    \normalfont B\kern-0.5em{\scshape i\kern-0.25em b}\kern-0.8em\TeX}}}

\acmConference[KDD '20]{KDD '20: ACM Humanitarian Workshop} {August 24, 2020}{San Diego, CA}

\begin{document}

\title{Identifying Meaningful Indirect Indicators of Migration for Different Conflicts}

\author{Lisa Singh, Katharine Donato, Ali Arab, Tomas Alvarez Belon, Abraham Fraifeld}
\authornote{Contact author: Lisa Singh - lisa.singh@georgetown.edu}
\author{Sean Fulmer, Douglas Post, Yanchen Wang}
\affiliation{%
  \institution{Georgetown University}
  \city{Washington}
  \state{DC}
  \postcode{20057}
}

\renewcommand{\shortauthors}{Singh et al.}

\begin{abstract}
This extended abstract describes an ongoing project that attempts to blend publicly available organic, real time behavioral data, event data, and traditional migration data to determine when and where people will move during times of instability. We present a methodology that was successful for a case study predicting mass movement in Iraq from 2015 - 2017, and discuss how we are extending it to capture indirect indicators of movement in Venezuela.  
\end{abstract}



\maketitle

\section{Introduction}
Forced migration is a worldwide crisis. According to UNHCR, one person is forcibly displaced every two seconds and the overall number of displaced persons continues to grow \cite{UNHCR2016}. Unfortunately, it is difficult to determine when mass displacement is going to occur and where people will move. 
Due to the lack of large-scale behavioral and movement data, local and national governments are frequently unaware of migration patterns until they occur. This deepens the inability of governments and aid organizations to effectively allocate resources, and provide safety for those that need it. 

A group of interdisciplinary researchers across multiple universities and institutions began working together to identify ways to combine variables constructed from publicly available organic forms of data with more traditional variables to forecast movement. Organic data refers to non-design data sources, including digital trace data. 
We were able to successfully integrate both publicly
available organic data from social media, specifically Twitter, and newspapers with more
traditional indicators of forced migration to determine movement patterns for a case study in Iraq \cite{Singh2020}. We combined movement and organic variables with spatial and temporal variation within different Hierarchical Bayesian models and showed the viability of our method. An important finding of the analysis was that incorporating open-source generated conversation buzz and event variables
maintains or improves predictive accuracy over traditional variables alone. Another important finding was that variables constructed using Arabic conversation had stronger predictive power than those constructed using English conversation. This is an indication of the importance of constructing variables from conversations in the languages that are most prevalent in the regions where displacement is occurring. 

Given the success of our approach for the Iraq case study, we are interested in determining if our approach can be used in other regions of the world and whether it can be adapted to new regions rapidly. This paper describes our general approach and explains our process for going from the initial Iraq case study to predicting movement in Venezuela. If our approach is accurate in more than one region of the world, we may be able to use real-time signals from different organic sources within our model to predict future movement across the world. This would help officials intervene, e.g. prepare aid or make resources available, before the largest movements occur. 

\section{General Approach}
The prediction task in Singh et al. \cite{Singh2020} was to estimate the number of families moving from one location to another. The high-level methodology used in that study was as follows:
\begin{itemize}
    \item \textbf{Identify relevant factors.} Migration researchers on the team have developed a dynamic factor model that shows the casual relationship between the different drivers of movement and the decision about whether or not to move \cite{MARTIN2019}.
    \item \textbf{Identify traditional and open-source data.} For this step, we find data for constructing variables that can serve as direct or indirect indicators of factors in the dynamic factor model. Survey data and movement tracking data can provide a lot of information about movement if they are available. 
    \item \textbf{Determine indirect indicators.} This step is critical when there are large gaps in traditional variables of movement. Here we identify conversation signals and events that are reasonable indirect indicators of different movement factors. For example, in the Iraq case study, discussion about violence in Iraq was a reasonable proxy (leading indicator) for death counts. 
    \item \textbf{Construct model.} There are many different options for this predictive task. Because we have spatial data and temporal data at different spatial scales and time resolutions, we need models that can handle this variation effectively.  The two most prevalent ones given our constraints are agent-based models and Hierarchical Bayesian models. 
    \item \textbf{Validate model.} One of the more challenging parts of this task is to find reliable ground truth data. These data are important for validating the prediction results. The International Organization for Migration (IOM) maintains a tracking matrix that contains movement information.\footnote{\url{https://www.iom.int/}} This is a reliable source for validation if data have been collected for the region of interest. If not, collecting survey data is the traditional approach for validating predictive models. 
\end{itemize}

The component of the methodology that we have focused on the most to date is identifying meaningful social media and newspaper signals. These forms of data give us an opportunity to learn about relevant events, topics people are discussing, and perceptions of the current situation. Our approach requires an understanding
of the importance and the changing dynamics of each migration factor/driver in a particular location. It is difficult to obtain these types of timely event, topic buzz, and perception data from traditional surveys or available administrative data. Therefore, we must consider alternate sources of data for constructing meaningful, dynamic variables that capture local changing conditions. We use social media and newspaper data to construct
event, buzz, and perception variables specific to each location (city or district) and each migration factor (economic, political, environmental, etc.), allowing them to serve as proxies or indirect indicators of the factors. For example, topic buzz about a flood can serve as an indirect indicator of a variable associated with the environmental factor. We can monitor the buzz to determine if it is constant, increasing, or decreasing through time.
Over the past few years, there have been enormous improvements in the accuracy of event, buzz, and perception (sentiment, stance, and emotion) detection methods in the context of migration from social media and newspaper data  \cite{AGRAWAL2016}, \cite{HOCKETT2018}, \cite{WEI2018}. While we only mention the most relevant methods, the advancement in this space has enabled us to capture meaningful signals through time in both Arabic and English.

\section{Forced Migration in Venezuela}

There are different ways we can expand our initial work. From a data mining perspective, the noise, bias, and misinformation in these domains reduces the accuracy of our current signal detection algorithms, meaning that we can improve algorithms for event detection, topic buzz, and perception by taking bias and misinformation into account. While we do want to calibrate poor quality information to understand the role it plays in influencing movement decisions, we do not plan on removing the misinformation because those who are considering moving may not realize it is biased or incorrect during their decision-making process.  From a machine learning perspective, because of the spatial and temporal variation, more robust models need to be created to effectively incorporate this variation. To date, we have focused on a systems dynamics approach using Hierarchical Bayesian models. We are investigating two additional alternatives. The first is extending existing Hierarchical Bayesian models to more elegantly handle these variations. The second is to use agents-based models to more extensively model different types of interactions among actors and also allow for  easier implementation of \textit{what if} scenario analysis. 

While we continue to work on these computer science and statistics problems, we also believe there is an urgent need to understand the value of these different signals for other regions of the world. One of the largest modern day humanitarian crises is taking place in Venezuela. According to UNHCR, between 2016 and the end of 2019, more than 4.6 million people left Venezuela. This is the largest displacement of people in recent history in Latin America. Migrants include people of working age, children and families, and both high and low skilled \cite{UNHCR2020}. 
Given the mass displacement in Venezuela and the different drivers of the conflict when compared to Iraq, we are extending our methods to improve our understanding of this crisis.  

\begin{figure*}[t]
  \centering
  \includegraphics[width=0.6\linewidth]{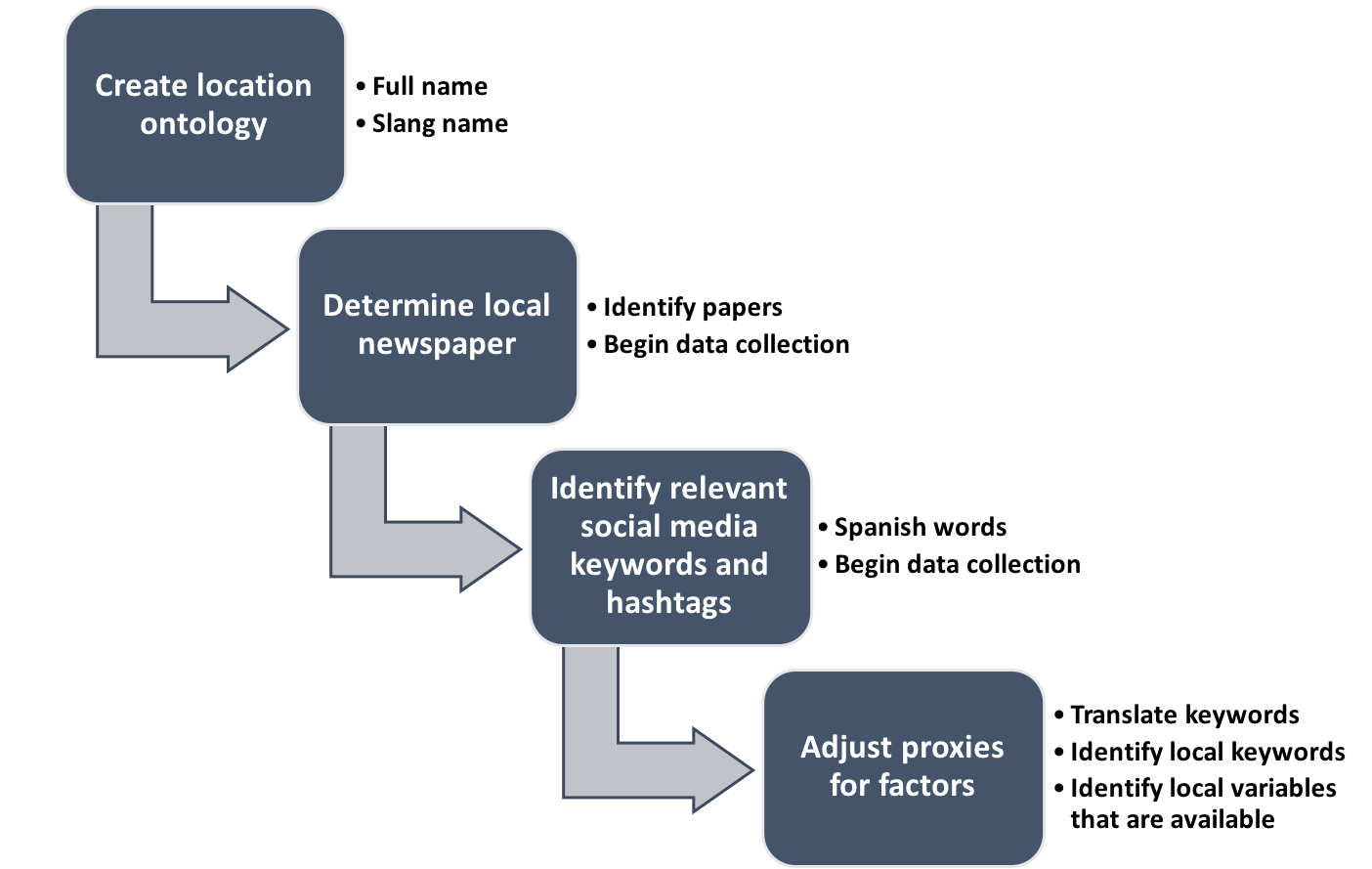}
  \caption{Methodology for capturing possible indirect indicators of forced migration in Venezuela}
\label{method}
\end{figure*}

Figure \ref{method} shows our approach for capturing possible indirect indicators of forced migration in Venezuela. While some steps can be automated, much of this is currently manually done, making this a costly process. We begin by creating a location ontology that identifies the cities, states, and regions of Venezuela and other neighboring countries. We use Statoids as the basis for this ontology.\footnote{\url{http://www.statoids.com}} We also identify both official and slang names since slang is common on social media. We then use the location ontology, existing databases of Venezuelan media, e.g. PrensaEscrita, and Wikipedia to identify newspapers across the region. We checked each source to ensure that the newspapers are legitimate. We also used Google keyword searches to further enhance our list. Local newspapers in the local language are generally more reliable than social media, and are particularly important for identifying local changing conditions and for detecting regional events. We are beginning data collection for these newspapers and are planning to rate the newspapers for bias and reliability. Many newspapers in Venezuela are government operated and may contain biases. Over time, we will want to measure the influence of these more biased sources. 

The next step involves identifying and collecting relevant keywords and hashtags from social media in the local language, Spanish. For each factor in the list of forced migration drivers, we identify words and hashtags that we believe are indicators of discussion about the factors of displacement, both generic determinants such as economic, social and environmental terms, as well as additional words that reflect the conditions of the individual crisis. For example, religious and ethnic tensions play a larger role in Iraq, while government and economic pressures are more significant in the migration decision-making process in Venezuela. Therefore, in the Iraq case study, including ISIS is important, but the extremist group is not relevant for Venezuela. Our team of research assistants identified Spanish words for each factor using the English and Arabic mappings developed for the Iraq case study. We also enriched the word list by identifying words that frequently co-occurred with the terms \textit{violence} and \textit{migration} in a sample of over 800 regional Spanish-language newspaper articles. Finally, we adjust the topics associated with each migration factor to include local phrases and words that are directly or indirectly associated with movement. A similar process was used for Portuguese since Brazil is a neighboring country of Venezuela and Portuguese is the primary language in Brazil.  We currently have approximately 850 words in Portuguese, 815 words in Spanish, and 740 words in Arabic. 

We have completed these steps for the Venezuela project and are beginning data collection. Once we have six months of data, we can build models and assess how well our models work in South America. We can also see if language in this case study plays as important a role as it did in the Iraq case study. 
In terms of ground truth data and validation, because movement data has not been extensively collected in Venezuela, team members at the Institute for the Study of International Migration will conduct a survey during the next year.

\section{Final Thoughts}
The paucity of disaggregated data in developing nations continues to be a stumbling block for detailed forced migration research. For example, due to the violence and instability persistent in Iraq, a nationwide census has not been conducted since 1997. This severely limits the ability for research into localized phenomena, at a time when it is most important. Furthermore, additional longitudinal data would help assess the change in these patterns. 

This methodology and project are focused on using available data in innovative, scalable ways to help identify mass displacement before it happens. It remains to be seen if organic data maintain their predictive power in the case of a humanitarian crisis in a separate region like South America. It could be the case that models constructed using social media and newspapers perform well in Iraq, but are less successful in other countries. Ultimately, it is important to organize access to data and algorithms that can help policy makers make informed decisions to help reduce levels of forced displacement. As the project continues, we will post updates, data, and code on our website \cite{VIZ}. We have a number of research partnerships in place and look forward to identifying new partners to increase our knowledge and capacity to help make progress on this societal-scale problem.

\section*{Acknowledgements}
This work was supported by the Massive Data Institute (MDI) at Georgetown University and the Institute for the Study of Migration (ISIM) at Georgetown University. 

\bibliographystyle{ACM-Reference-Format}
\bibliography{refs}

\end{document}